# Extração e Classificação de Características Radiômicas em Gliomas de Baixo Grau para Análise da Codeleção 1p/19q

T.A.M. Silva[1,2], G.S. Cassia[3], J.L.A. Carvalho[1]

[1] Departamento de Engenharia Elétrica, Faculdade de Tecnologia, Universidade de Brasília, Brasília-DF, Brasil
[2] Instituto Federal Goiano - Campus Cristalina, Cristalina-GO, Brasil
[3] Mezo Diagnósticos, Rede D'Or São Luiz Regional Brasília/Hospital Daher, Brasília-DF, Brasil

*Resumo* — **A radiômica é uma área emergente, que apresenta um grande conjunto de métodos e técnicas computacionais para extrair características quantitativas de imagens de ressonância magnética. Na etapa de extração de características, suas saídas devem ser bem definidas e cuidadosamente avaliadas, para fornecer diagnósticos por imagem, prognósticos e respostas ao tratamento de terapias. Neste estudo, apresentamos a extração de características quantitativas a partir de imagens de ressonância magnética em gliomas de baixo grau utilizando a biblioteca *Pyradiomics* e, usando uma rede neural perceptron multicamadas, mostraremos a predição da deleção dos cromossomos 1p/19q nesses gliomas. Vários estudos mostram que a codeleção cromossômica 1p/19q é um fator prognóstico positivo nos gliomas de baixo grau, por serem mais sensíveis ao tratamento quimioterápico. Devido ao grande número de características extraídas, necessitou-se utilizar de uma técnica de redução de dimensionalidade, a análise de componentes principais, que se mostrou eficiente neste estudo. Após o treinamento e teste das características realizados pela rede neural perceptron multicamadas os resultados mostraram-se bastante promissores na detecção do status de deleção dos cromossomos 1p/19q, principalmente levando em consideração a possibilidade de se evitar biópsias cirúrgicas para este diagnóstico.**

*Palavras-chave* — **Radiômica, extração de características quantitativas, rede neural perceptron multicamadas, *Pyradiomics*, análise de componentes principais.**

## I. Introdução

A ressonância magnética (RM) é um método de imagem médica não invasiva, fornecendo excelente contraste de tecidos moles e é reconhecidamente o método padrão-ouro para o diagnóstico e controle de tratamento de tumores cerebrais [1, 2]. A radiômica pode ser utilizada para extração de parâmetros quantitativos de imagens de RM, fornecendo dados calculados mineráveis e não visuais, e, consequentemente tornando-os importantes no apoio à decisão clínica [3] e previsão de prognóstico de tumores intracranianos e classificação pré-cirurgica [4,5]. Nesse contexto, os dados quantitativos extraídos por análise radiômica podem refletir características histopatológicas e genéticas de tumores.

A Organização Mundial da Saúde (OMS) classifica os tumores cerebrais de acordo com características histológicas que refletem seu nível de agressividade. Os gliomas de baixo grau (*low-grade gliomas*, ou LGG) são considerados tumores de menor agressividade e com melhor prognóstico quando comparados aos gliomas de alto grau (*high-grade gliomas*, ou HGG), como o glioblastoma (OMS grau IV) [6]. Neste artigo, propomos a possibilidade de uma classificação não invasiva de gliomas de baixo grau, com base na codeleção cromossômica 1p/19q [3,7].

Em sua mais recente atualização a OMS recomenda uma classificação integrativa, acrescentando marcadores moleculares e genéticos às características histológicas dos tumores, permitindo uma melhor adequação da terapia de pacientes, melhor classificação para ensaios clínicos e estudos experimentais e uma categorização mais precisa [8]. Dessa forma, atualmente os tumores do tipo oligodendroglioma são caracterizados pela codeleção dos cromossomos 1p/19q, enquanto que os astrocitomas não apresentam essa codeleção cromossômica. Alguns estudos mostraram que a codeleção dos cromossomos 1p/19q deve ser considerada como importante marcador molecular tumoral preditivo de resposta ao tratamento quimioradioterápico e maior sobrevida, em comparação aos tumores não-codeletados [9-11]. Outros estudos realizaram a extração de características radiômicas para mostrar o status de codeleção cromossômica 1p/19q. Por exemplo, Akkus et al. [12] extraíram dados das imagens a partir da varredura das regiões de interesse (já segmentadas por médicos radiologistas) pela rede neural convolucional. Já no estudo conduzido por Zhang et al. [13], dois radiologistas realizaram a segmentação manual das regiões de interesse em cortes axiais, fatia a fatia, posteriormente validadas por outros dois radiologistas seniores, e então foi aplicada a ferramenta LIFEx Soft [14] para extração de 40 características quantitativas de primeira e segunda ordem. Estudos recentes demonstram uma abordagem mais precisa e reprodutível para captura de informações computacionais e posterior extração de caraterísticas de forma e texturas tumorais em imagens de RM de crânio [15,16,17].

Apesar da extração de características ser uma etapa de fundamental importância dentro da radiômica, muitas vezes a quantidade de características extraídas torna complexo o processamento das informações nas etapas posteriores do mo-delo projetado. Sendo assim, considera-se a redução da dimensionalidade uma operação fundamental para viabilizar a visualização de dados multidimensionais, baseando-se em transformações sobre os dados projetando-os em espaços de menor dimensão, com máxima relação de vizinhança entre os dados [18]. Em 1997, foi publicado um estudo [19] que comparou métodos de redução de dimensionalidade, incluindo a análise de componentes principais (*principal component analysis*, ou PCA) aplicada à classificação de padrões. Os resultados foram aferidos de forma indireta, pela taxa de acerto da classificação binária de amostras de uma análise citológica. Uma década depois, Yin [20] mostrou uma avaliação geral dos seis principais métodos de redução de dimensionalidade e realizou uma comparação entre eles. Uma das técnicas avaliadas foi a PCA. Os experimentos constituíam-se em projetar de três para duas dimensões um conjunto de dados artificiais, em que os dados tridimensionais de entrada estavam distribuídos uniformemente e as comparações basearam-se em visualizações gráficas dos dados bidimensionais de saída.

De posse das características extraídas e redimensionadas, em nosso estudo uma rede neural do tipo perceptron multicamadas (*multilayer perceptron*, ou MLP) foi treinada e testada com estas informações. Ao longo do tempo, redes MLP têm apresentado sucesso em várias aplicações importantes, como classificação de padrões, interpolação de funções, otimização, predição e controle [21]. Ciresan et al. [22] apresentaram um estudo sobre classificação de dígitos manuscritos, apontando que os melhores resultados obtidos na competição referem-se à utilização de um número maior de camadas ocultas de neurônios e de dados para evitar *overfiting*. O estudo relata a utilização de poderosa placa de processamento gráfico para obter ganho no desempenho computacional.

Em nosso estudo, apresentamos um método simples, promissor e não invasivo para predizer o status de codeleção cromossômica 1p/19q de LGG a partir de imagens de RM ponderadas em T2 usando redes neurais MLP.

## II. Material e métodos

Para este estudo, utilizamos o *dataset* LGG-1P19QDeletion do The Cancer Imaging Archive [23]. Esse conjunto de dados apresenta imagens de RM pré-operatória de 159 pacientes com LGG, mapeados em um período de 9 anos. Todos os pacientes tiveram biópsia comprovando o status dos cromossomos 1p/19q (codeletados ou não-codeletados). As imagens foram coletadas na Mayo Clinic, Minnesota, USA, administrado por Mayo Foundation for Medical Education and Research. São 102 pacientes com status de codeleção e 57 pacientes com status de não-codeleção dos cromossomos 1p/19q. Destes LGGs, 17 são astrocitoma, 45 são oligodendroglioma e 97 oligoastrocitoma. As imagens foram adquiridas na ponderação T2 e classificadas em tumores 1p/19q codeletados e 1p/19q não-codeletados, conforme exemplo na figura 1. Segundo Akkus et al. [12], as imagens foram adquiridas utilizando um protocolo muito consistente, incluindo cortes de 1 mm e 3 mm em aparelhos de 1,5 T e 3 T. As regiões de interesse foram segmentadas em duas dimensões, utilizando ferramenta semiautomática desenvolvida pelos mesmos.

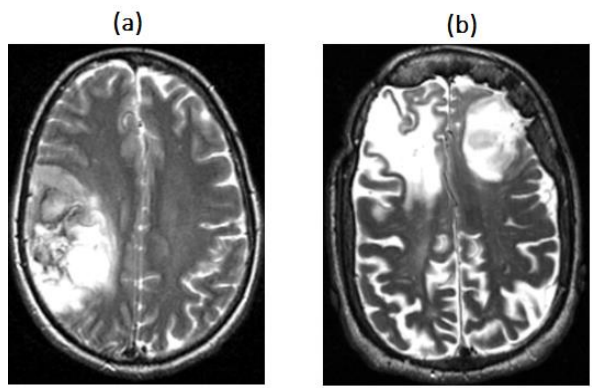

Fig. 1 – Exemplos de imagens de RM de gliomas de baixo grau (LGG): (a) cromossomo 1p/19q codeletado; e (b) cromossomo 1p/19q não deletado.

A preparação dos dados para treinamento da rede neural será realizada em quatro etapas: extração, normalização, redimensionamento e aumento dos dados. Após estas etapas, os dados serão utilizados para o treinamento na MLP a fim de predizer o status dos cromossomos 1p/19q (codeleção ou não).

### A. Pré-processamento

*Extração de características:* Após *download* e descompactação do *dataset*, utilizamos a ferramenta 3DSlicer [25] para realizar a conversão das imagens de RM do formato .nii para .nrrd. Em seguida, utilizou-se a biblioteca Pyradiomics para extração das características quantitativas. O pacote Pyradiomics [26] é uma plataforma flexível de código aberto para extração de grande número de características quantitativas em imagens médicas, implementada em linguagem de programação Python. A plataforma Pyradiomics usa quatro etapas principais na extração de características radiômicas das imagens: (i) carregamento, pré-processamento e mapas de segmentação da imagem; (ii) aplicação de filtros; (iii) cálculo das características dentre as classes propostas; e (iv) retorno dos resultados. A figura 2 ilustra esse processo.

Neste estudo, foram extraídas um total de 120 características conforme mostra a tabela 1.

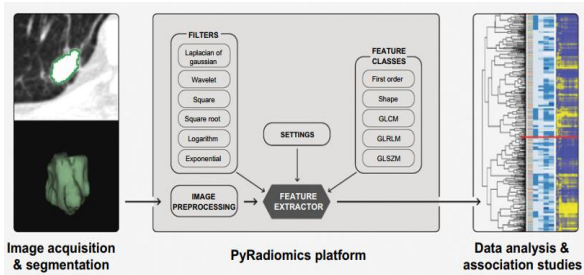

Fig. 2 – Etapas da plataforma PyRadiomics [26] (reproduzida com permissão).

Tabela 1 – Quantitativo de características extraídas.

| classe de características [27] | quantidade de características |
|---|---|
| estatísticas de primeira ordem | 19 |
| baseado em forma (3D) | 16 |
| baseado em forma (2D) | 10 |
| matriz de coocorrência de nível de cinza | 24 |
| matriz de duração de nível de cinza | 16 |
| matriz de tamanho de zona de nível de cinza | 16 |
| matriz de dependência de nível de cinza | 14 |
| matriz de diferença de tons de cinza vizinhos | 5 |

Além da extração das características, a plataforma Pyradiomics permite também que filtros possam ser aplicados na imagem original [27]. Em nosso estudo, foram utilizadas características originais, ou seja, extraídas sem aplicação de nenhum filtro.

*Normalização de dados:* As características extraídas através da plataforma Pyradiomics foram então preparadas para obtenção de melhor resultado na etapa de classificação. A primeira ação realizada foi a de normalização. Essa etapa é importante para desempenho de algoritmos de aprendizado de máquina. Sem a normalização, percebemos que a acurácia na etapa de classificação pela MLP sempre estava entre 60% e 65%. A normalização também é importante tendo em vista o uso da PCA, que é efetuada por escala. Portanto, utilizamos do método *StandardScaler* da biblioteca *scikit-learn* do Python para realizar tal tarefa. O método *StandardScaler* padroniza o conjunto de dados de modo que estes tenham média nula e desvio padrão unitário.

*Redimensionamento dos dados:* Devido ao grande número de características de entrada (tabela 1), aplicamos a técnica de PCA logo após a normalização dos dados. A PCA é uma técnica de redução de dimensionalidade linear que pode ser utilizada para extrair informações de um espaço de alta dimensão, projetando-as em um subespaço de menor dimensão [28]. A PCA tenta preservar as partes essenciais (que têm mais variação dos dados) e remover as partes não essenciais (com menos variação). Podemos aproveitar esta característica para acelerar os algoritmos de aprendizado de máquina.

Neste trabalho, redimensionamos as 120 colunas de dados de entrada em 8 colunas. As 8 componentes apresentaram os seguintes percentuais de redução: PCA1=47, PCA2=19, PCA3=13, PCA4=5, PCA5=3, PCA6=3, PCA7=2 E PCA8=2, totalizando 92% do redimensionamento. A figura 3 traz uma apresentação visual desse resultado.

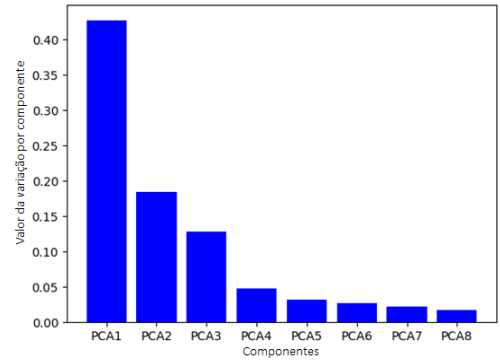

Fig. 3 – Resultado do dimensionamento dos dados de entrada.

*Geração de dados sintéticos:* Akkus et al. [12], que também estudaram a codeleção cromossômica 1p/19q, redimensionaram sua base de dados para obter um maior número de casos para treinamento e teste. Em nosso trabalho, utilizamos o método SMOTE (*synthetic minority oversampling technique*, ou técnica de sobreamostragem sintética de minorias). O método SMOTE equilibra a distribuição de classes, aumentando aleatoriamente exemplos de classes minoritárias. Ele gera os registros de treinamento por interpolação linear para a classe minoritária. Esses registros de treinamento sintético são gerados ao selecionar aleatoriamente um ou mais *k* vizinhos mais próximos para cada exemplo na classe minoritária [29]. Em nosso *dataset* de 159 casos, ao separarmos base de treinamento e teste (75% e 25% respectivamente), temos na classe de saída 119 casos para treinamento; destes casos, 76 estavam classificados como "0" (oligodendroglioma) e 43 casos como "1" (astrocitoma). A partir da classe minoritária, o método SMOTE gerou mais 33 casos sintéticos para o balanceamento dos casos de saída, obtendo assim 76 casos de cada rótulo. Assim, obteve-se uma matriz de 152 linhas (casos para treinamento) e 8 colunas (características redimensionadas pela PCA).

### B. Redes neurais perceptron multicamadas

As redes neurais artificiais (RNAs) são consideradas uma das principais ferramentas para reconhecimento de padrões, estando bastante consolidadas no meio científico e computacional [30]. Dentre as RNAs mais difundidas encontram-se as MLPs, que têm sido utilizadas em detecção de doença dos olhos, predição de riscos de terremotos, reconhecimento de face, classificação de sinais de eletromiografia, dentre outros [31]. Contudo, é uma técnica de reconhecimento de padrões com uma fase de treinamento demorada e de grande esforço computacional [32]. A figura 4 ilustra a arquitetura de uma RNA MLP.

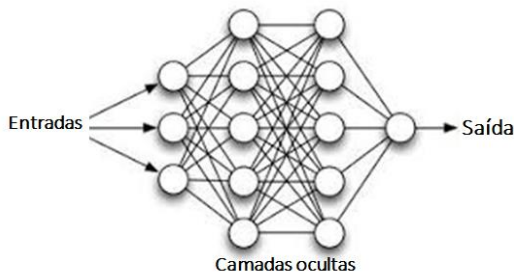

Fig. 4 – Arquitetura de uma rede MLP (exemplo ilustrativo).

A rede MLP utilizada neste trabalho caracteriza-se por ser uma rede sequencial, com uma camada de entrada, duas camadas ocultas e uma camada de saída. Utilizamos características de uma rede densa (conhecida como *full connect*), em que todos os neurônios se conectam aos neurônios da camada subsequente. A camada de entrada recebe dados após a etapa de pré-processamento. Nas camadas ocultas, todos os neurônios possuem a mesma dimensão e aplicam uma função de ativação a cada neurônio da camada anterior. A função de ativação utilizada nestas camadas é a RELU (unidade linear retificada), definida como $f(x)=\max(0,x)$, que é atualmente uma das funções mais amplamente usadas [33]. A função de ativação RELU não ativa todos os neurônios ao mesmo tempo, ou seja, se o valor de entrada for negativo, será atribuído zero a ele e o neurônio não é ativado. Na camada de saída utilizamos a função de ativação sigmoide, também amplamente conhecida, e definida como $f(x)=1/(1+e^{-x})$, com valores de $f(x)$ variando entre 0 e 1 [34].

### C. Implementação

Iniciamos nossa implementação da MLP utilizando a biblioteca PyBrain, mas a mesma apresentava poucos recursos de otimização. Em seguida migramos para a biblioteca Keras, pois aproveita muito bem as unidades de processamento gráfico e acelera os algoritmos de aprendizado de máquina [35], além de implementar uma infinidade de métodos para RNAs e dispor de extensa documentação. Em nossa camada de entrada, usamos 8 atributos previsores (ou seja, as características obtidas após dimensionamento pelo PCA) e 192 casos (obtidos após etapa de geração de dados sintéticos através do SMOTE). Nas camadas ocultas, utilizamos 5 neurônios (metade da camada de entrada mais 1) e função de ativação *rectified linear unit* (ReLU). Na inicialização dos pesos, utilizamos a função *random_uniform*, que inicializa os pesos através de uma distribuição uniforme entre −0,05 e 0,05. Como nosso trabalho busca a informação de deleção ou não deleção do cromossomo 1p/19q, temos uma resposta do tipo binária; portanto, utilizaremos apenas um neurônio na camada de saída e a função de ativação será a sigmoide.

Ao compilar a MLP, um dos parâmetros utilizados é o *adam*, que ativa uma função de ajuste dos pesos que usa otimização por descida de gradiente estocástica [34, 35]. Como estamos trabalhando com dados com muitas dimensões, essa função é uma das mais indicadas e o que melhor se adapta ao tipo de trabalho proposto. Outro parâmetro utilizado foi o *binary_crossentropy* [35], que ativa uma função de tratamento do cálculo do erro para problemas de classificação binária. Por fim, a métrica utilizada para fazer a avaliação da acurácia foi a *binary_accuracy* [35], justamente por estarmos trabalhando com problema de somente duas classes.

Como se trata de um aprendizado supervisionado, ao executar o treinamento da rede, a MLP vai relacionar os atributos previsores e os atributos de classe de saída conforme dividido anteriormente em 75% e 25%, respectivamente. Juntamente com a execução da MLP passamos o parâmetro *batch_size* = 10, que realiza o ajuste dos pesos a cada dez registros, o que pode fazer bastante diferença nos resultados. Outro parâmetro utilizado na execução da rede é o *epochs*, que define quantas vezes realizamos o treinamento (quantas vezes realizamos o ajuste dos pesos). Conforme tabela 2, executamos o parâmetro com 100 e 1000. Muito provavelmente, a cada execução os valores não serão idênticos, pois os pesos são ajustados aleatoriamente a cada execução.

Como já descrito na seção II-A, após aumento sintético dos dados e separação da base de dados em treinamento e teste, ficamos com 152 casos (75%) para treinamento e 40 casos (25%) para teste.

A configuração dos parâmetros acima escolhidos mostrou-se aquela com melhor desempenho. Para esta configuração, foi calculada a sensibilidade, especificidade e acurácia.

A sensibilidade é calculada pela taxa de positivos verdadeiros (*true positives rate*, ou TPR), sendo TPR = TP / (TP+FN), em que TP são os positivos verdadeiros e FN são os falsos negativos. A especificidade é calculada pela taxa de negativos verdadeiros (*true negatives rate*, ou TNR), sendo TNR = TN / (TN+FP), em que TN são os negativos verdadeiros. A acurácia (ACC) é calculada como ACC = (TP + TN) / (TP + FP + TN + FN).

Também comparamos os resultados da classificação da MLP com outros dois algoritmos clássicos de classificação, o *support vector machine* (SVM) e o *random forest* (RF).

## III. Resultados

A tabela 2 mostra o desempenho da rede MLP na etapa de treinamento. Observamos que, ao longo das épocas de treinamento, o algoritmo inicia com um valor ruim de acurácia e taxa de perda, mas vai se adaptando até obter um bom resultado final. Com 100 épocas, a rede consegue alcançar acurácia de 100% e taxa de perda de 0%.

Tabela 2 – Execução da MLP no treinamento.

| épocas | acurácia | | taxa de perda | |
|---|---|---|---|---|
| | início | fim | início | fim |
| 10 | 67,0% | 94,0% | 1,2% | 0,0% |
| 100 | 89,0% | 100,0% | 0,8% | 0,0% |

A tabela 3 apresenta os resultados de precisão, especificidade e sensibilidade na base de validação, tanto para a MLP, como para a SVM e a RF. O algoritmo da MLP foi testado com 100 épocas. Precisão, especificidade e sensibilidade são as principais características de performance para avalição de algoritmos de classificação. O método proposto apresentou a melhor precisão (85%) e sensibilidade (88%) dentre os algoritmos avaliados; com relação à sensibilidade (73,5%), o algoritmo proposto apresenta resultado melhor que a SVM (70%) e resultado muito próximo ao apresentado pela RF (75%). Dentre os três resultados, destacamos o de precisão, que representa a acurácia do modelo no momento de validação: o algoritmo proposto apresentou a melhor acurácia dentre os três algoritmos testados, com 85% de probabilidade de fornecer resultado correto.

Tabela 3 – Execução da MLP, SVM e RF na base de validação.

| Algoritmo | precisão | especificidade | sensibilidade |
|---|---|---|---|
| MLP | 85,0% | 73,5% | 88,0% |
| SVM | 77,5% | 70,0% | 85,0% |
| RF | 75,0% | 75,0% | 75,0% |

A tabela 4 mostra a matriz de confusão do status 1p/19q na base de testes após execução do algoritmo da MLP. Em um total de 40 casos, a rede errou apenas 3 casos de deleção (falsos negativos) e 4 casos de não deleção (falsos positivos). Sendo assim, acertou 22 casos que são positivos verdadeiros (casos de deleção) e 11 casos que são negativos verdadeiros (casos de não deleção).

Tabela 4 – Matriz de confusão do status 1p/19q na base de testes.

| casos de teste = 40 | codeletado (real) | não-codeletado (real) |
|---|---|---|
| codeletado (predito) | 22 | 4 |
| Não-codeletado (predito) | 3 | 11 |

## IV. Discussão

Um grande desafio para execução de algoritmos de aprendizado de máquina é a obtenção de um número adequado de imagens para treinamento e testes. A execução da MLP mostrou capacidade de aprender as características extraídas da imagem, obtendo melhor acurácia na base de treinamento sobre a base de testes, sobretudo pela quantidade de casos ser pequena na base de testes. Conforme resultados da tabela 3, a MLP apresentou uma precisão melhor que os demais métodos comparados.

Vale registrar aqui que as 120 características, quando utilizadas sem nenhum pré-processamento para treinamento da MLP, resultaram em um valor de acurácia de 65% na etapa de treinamento. A normalização, o redimensionamento e o aumento na base de dados foram importantes para um resultado mais robusto da MLP sobre as demais técnicas usadas, conforme mostra a tabela 3. Os valores de especificidade e sensibilidade obtidos indicam que a MLP foi capaz de identificar com bom percentual de precisão as características de codeleção cromossômica em gliomas de baixo grau.

## V. Conclusão

Neste trabalho foi apresentado uma técnica não invasiva baseada em MLP que se mostrou eficaz para predizer a codeleção cromossômica 1p/19q através de análise de características extraídas de imagens de RM ponderadas em T2. A identificação da codeleção cromossômica 1p/19q em pacientes com LGG representa uma oportunidade positiva de se determinar melhores decisões terapêuticas. Essa tomada de decisão é importante, pois tumores codeletados (oligodendrogliomas) tem melhor resposta quimioterápica e, portanto estes pacientes tem uma maior sobrevida global, em comparação aos pacientes com tumores não-codeletados (astrocitomas).

## Conflitos de Interesse

Os autores declaram que não tem conflitos de interesses.

Autor: Tony Alexandre Medeiros da Silva
Instituto: UnB – Universidade de Brasília
Endereço: Campus Universitário Darcy Ribeiro, Asa Norte
Cidade: Brasília-DF
País: Brasil
Email: tonyufu2005@gmail.com